\documentclass[a4paper]{spie}  

\usepackage{graphicx} 
\usepackage{amsmath,amsfonts,amssymb,mathrsfs}

\title{Optimising the multiplex factor of the frequency domain multiplexed readout of the TES-based microcalorimeter imaging array for the X-IFU instrument on the Athena Xray observatory.}

\author[a]{J. van der Kuur}
\author[a]{L. Gottardi}
\author[a]{H. Akamatsu}
\author[a]{B.J. van Leeuwen}
\author[a]{R. den Hartog}
\author[a]{D. Haas}
\author[c]{M. Kiviranta}
\author[b]{B.J. Jackson}
\affil[a]{SRON Netherlands Institute for Space Research, Sorbonnelaan 2, 3584CA Utrecht, the Netherlands}
\affil[b]{SRON Netherlands Institute for Space Research, Landleven 12, 9747AD Groningen, the Netherlands}
\affil[c]{VTT, Tietotie 3, FIN-02150 Espoo, Finland}

\authorinfo{Corresponding author: J. van der Kuur\\E-mail: j.van.der.kuur@sron.nl, Telephone: +31 88 777 5862\\ Accepted for publication in the proceedings of the SPIE, doi 10.1117/12.2232830.\\The copyright has been transferred to the SPIE.}

\begin{document} 
\maketitle

\begin{abstract}

Athena is a space-based X-ray observatory intended for exploration of the hot and energetic universe.
One of the science instruments on Athena will be the X-ray Integrated Field Unit (X-IFU), which is  a cryogenic X-ray spectrometer, based on a large cryogenic imaging array of Transition Edge Sensors (TES) based microcalorimeters operating at a temperature of 100mK. The imaging array consists of ±3800 pixels providing 2.5 eV spectral resolution, and covers a field of view with a diameter of of 5 arc minutes. 

Multiplexed readout of the cryogenic microcalorimeter array is essential to comply with the cooling power and complexity constraints on a space craft. Frequency domain multiplexing has been under development for the readout of TES-based detectors for this purpose, not only for the X-IFU detector arrays but also for TES-based bolometer arrays for the Safari instrument of the Japanese SPICA observatory. 

This paper discusses the design considerations which are applicable to optimise the multiplex factor within the boundary conditions as set by the space craft. More specifically, the interplay between the science requirements such as pixel dynamic range, pixel speed, and cross talk, and the space craft requirements such as the power dissipation budget, available bandwidth, and electromagnetic compatibility will be discussed. 

\end{abstract}


\section{INTRODUCTION}

Athena is a space-based X-ray telescope under development by ESA \cite{barcons15} to study the hot and energetic universe. It has been selected as the L2 mission in ESA's Cosmic Vision program, and as such has a projected launch date in 2028. One of the two scientific instruments is the X-ray Integral Field Unit (X-IFU), which is tailored for high resolution spectroscopy in the energy range between 0.2 -- 12~keV in a field of view of 5'. 

In the focal plane of the X-IFU instrument there will be an imaging array of micro-calorimeters, with an unprecedented energy resolving power of $\Delta E < 2.5$~eV @ 7~keV in combination with response times compatible with the observation of a 1~mCrab point source @ 80\% throughput efficiency. To achieve this performance, transition edge sensor (TES) based micro-calorimeter arrays operated at $\sim 90$~mK are the baselined detectors. 

A TES exploits the resistive transition of a superconductor to obtain a thermometer with a very high sensitivity, typically a factor of 10 -- 100 better than in semi-conductor based thermometers. This property makes it possible to build a detector array which meets the Athena requirements.

The electrical readout of this type of detector is challenging because of the low operating temperatures. Because cooling power is expensive, especially in resource limited space crafts, careful optimisation of the readout chain is essential to make the instrument feasible. The multiplexing factor is one of the key elements in this optimisation, as well as the power dissipation of the cryogenic SQUID amplifiers which are generally used to readout TES-based detectors.

This paper focusses on the design of the cryogenic part of the readout chain, as it requires a significant fraction of the required resources. The choices made for the baseline design will be motivated, and some options for further optimisation will be discussed.

\section{MULTIPLEXED READOUT OF TES-BASED MICROCALORIMETERS}

Readout of TES-based micro-calorimeters can be seen as monitoring the resistance of a temperature dependent resistor, which value is a function of the applied photon energy.  The monitoring is generally done by applying a constant voltage, and reading the temperature-dependent current, as voltage biasing creates negative electro-thermal feedback\cite{irwin95}. The negative feedback enhances the linearity and stability of the micro-calorimeter. 

To comply with the required voltage bias condition, a current or transimpedance amplifier is needed with an effective input resistance which is significantly smaller ($\lesssim 0.3 R_0$) than the set point resistance ($R_0=1.5\mathrm{m} \Omega$) of the micro-calorimeter. In addition to that, electro-thermal stability requires that the electrical inertia of  the bias circuit should be approximately a factor of 3 - 6 smaller than the thermal inertia. This limits the maximum input inductive reactance of the current amplifier at the thermal cutoff frequency of the micro-calorimeter to $\sim 1/6$ -- $1/3 R_0$.

Because of the stringent science requirements, it has been chosen that the readout circuit is allowed to consume $\lesssim 2.7$\% of the energy resolution budget. This implies that the required effective noise temperature of the current or trans-impedance amplifier should be no more than 20\% of the effective noise temperature of the micro-calorimeter, taking into account that the detector noise and the amplifier noise are uncorrelated. The latter noise temperature is approximately equal to the operating temperature, i.e. ~100mK. Hence, the effective noise temperature of the amplifier has to be smaller than $\sim$20mK. This requirement, combined with the low input impedance requirement, rules out the use of semiconductor amplifiers and leaves SQUID amplifier readout, either in a current or trans-impedance configuration, as the only viable option.

Multiplexed readout implies that multiple pixels are readout by a single SQUID amplifier, and multiple TES bias voltages are fed through one wire pair. Multiplexed readout as opposed to direct readout of TES-based cryogenic micro-calorimeters is mainly driven by the goal to minimise the required cooling capacity and complexity at the cold stages. The main sources of heat load in the readout system are the SQUID amplifiers, the heat leak through the bias and readout wiring, and the size (and mass) of the interconnections. Without any multiplexing, each pixel would require a single wire pair for its voltage bias, and three wire pairs for operating the SQUID amplifier, i.e. $\sim 16$k pairs for the X-IFU instrument. 

Simple addition of the TES signal currents in a single SQUID is not acceptable, because as the signals overlap in time, frequency and phase space,  they become indistinguishable after simple addition. 

The stringent Athena science requirements do not allow for significant energy resolution as a result of multiplexing, as this can be avoided. In text books it can readily be found that any lossless multiplexing scheme, i.e. a scheme in which the constituent signals need to remain fully recoverable, requires two fundamental steps before the signals can be added while remaining (mathematically) independent. The first step is to confine the signal bandwidth, and the second step is to transform the signals (coding) in such a way that they become mathematically independent. The latter mathematical transformation implies the multiplication of the time-dependent signal functions of the pixels with an independent carrier function per pixel, in such a way that they become orthogonal and therefore mathematical independent and fully distinguishable after addition (``multiplexing'') and amplification with a single SQUID amplifier. 

As a result of the mathematical transformation, $M$ multiplexed signals require a bandwidth which is at least $M$ times the bandwidth of a single signal, depending on the chosen set of carrier functions. As will become clear below, SQUID amplifiers can provide sufficient power gain over bandwidths of several MHz. On the other hand, X-IFU micro-calorimeter pixels require a bandwidth of less than 5 - 10 kHz per pixel (SSB). Therefore, from bandwidth point of view, SQUID amplifiers are in principle capable of accommodating the signals of at least several tens of  micro-calorimeter pixels.

There are many orthogonal (carrier) basis sets for multiplexing conceivable, but in practice only a few are commonly applied. For micro-calorimeter readout time domain (TDM), frequency domain (FDM), and code domain multiplexing (CDM) are actively being developed\cite{stiehl12,doriese16}. In TDM and CDM switching SQUIDs are used code the signals with boxcar and Walsh functions, respectively, while in FDM the each TES codes its signal by amplitude modulating a sinusoidal bias current of a distinct frequency per pixel. 

When optimally dimensioned, and with full signal loading, the multiplexed signals require the same resources, independent of the applied basis set. The proÕs and conÕs of each set merely depend on practical implementation issues.  
However, in the X-IFU case, where most the photons have a relative low energy because of the energy dependence of the effective area of the telescope ($\mathrm{area} \propto E^{-2}$), the dynamic range consumption in the SQUID amplifier in case of FDM is the smallest, as addition of detector responses to signals larger than the signal of a saturating photon is very small. As a result, tens of pixels can be readout with a single SQUID amplifier, without needing more dynamic range than what is needed for the readout of a single pixel.  Each coding scheme has its own set of implementation characteristics, but further discussion is outside the scope of this paper. FDM has been chosen as the baseline multiplexing scheme for X-IFU, with TDM readout as a backup solution.

\subsection{Implementation aspects of frequency domain multiplexing}

As already mentioned above, multiplexing involves mounting a bandwidth-limited signal on a carrier function by means of multiplication (modulation), followed by summing of the modulated signals for transportation through a shared channel. In the X-IFU frequency domain multiplexing (FDM) implementation, the orthogonal set of carrier functions consist of sinusoidal TES bias voltages of different frequencies. The signal-dependent TES resistance amplitude-modulates the amplitude of the bias current. 

\begin{figure}
\centering
\includegraphics[width=0.5\textwidth]{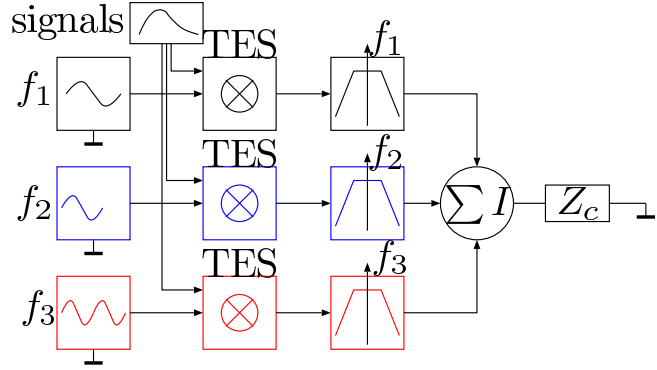}
\caption{Schematic FDM implementation scheme for a single column. Each TES in a column is biased with a sinusoidal voltage bias of a different frequency. The TESs amplitude modulate the resulting bias currents. The signals are separated in frequency space by means of tuned $LC$ bandpass filters, consisting of a coil ($L$) and capacitor ($C$). The filtered signals are summed in a current summing point, with finite impedance $Z_c$, which schematically represents the input impedance of the SQUID current amplifier. The SQUID amplifier chain has not been drawn in this schematic.}
\label{fig_FDMschematic}
\end{figure}

The required bandwidth limitation, and separation in frequency space is provided by $LC$ bandpass filters, consisting of an inductor ($L$) and a capacitor ($C$). A schematic diagram is shown in Fig.~\ref{fig_FDMschematic}. The bandwidth-limited signals are summed in a summing point, and measured by a SQUID with a total input impedance $Z_c$. Note that the rest of the SQUID amplifier chain is not shown in this diagram. 

The bandpass filters are connected in series with the TESs. As a result, their parasitic series resistances add to the internal resistance of the TES bias circuits. This internal resistance is limited by the voltage bias condition requirement to $\lesssim 0.3 R_0$, as mentioned earlier.  This implies that the intrinsic $Q$-factor of the $LC$ bandpass filters including the internal resistance of the voltage source has to be $\gtrsim 4$ times higher than the $Q$-factor or the circuit with the TES in transition, to satisfy the voltage bias condition.

\subsubsection{Electrical cross talk and EMI}

Electrical cross talk occurs when a pixel modulates more than one carrier, when modulated signals cross couple in the wiring harness between the detector and the room temperature electronics, or as a result of nonlinearity which creates mixing between signals.  

The first mechanism results from the fact that the attenuation provided by the bandpass filters is finite, so that cross talk occurs with an amplitude depending on the ratio of the bandwidth per pixel, and the frequency spacing between pixels.  There are three cross talk mechanisms  related to the $LC$ filters \cite{kuur03, dobbs12}. 

First, the finite summing point impedance (in practice this is an inductance) is a shared element between bias circuits, so that it becomes a medium to transfer energy between pixels. This effect scales as $Q_{pixel} L_c/L$,  where $Q_{pixel}$ is the ratio of the bandwidth and the carrier frequency of a pixel, and $L_c/L$ is the ratio of the common inductance and the filter coil inductance. The size of this effect sets an upper limit to the geometrical common inductance, and sets a lower limit to the energy resolution of the SQUID.

Second, magnetic cross-coupling between the inductances of the filters create a transformer coupling between bias circuits. This effect scales as $k Q_{sep}/2$, where $k$ is the coupling factor between the filter inductors, and $Q_{sep}$ the ratio of the frequency separation between two carriers, and the lower carrier frequency. The size of this effect depends on the proximity between filter coils, and can be further reduced by using gradiometric configurations.

Third, the multiplexed bias voltage comb creates finite leakage of neighbouring bias currents proportional to $(Q_{sep}/Q_{pixel})^2$. Note that the size of this effect can be traded against wire usage, by decreasing $Q_{sep}$ by splitting the bias comb over multiple wire pairs.

Cross talk in the wiring harness can be lowered by a combination of a careful symmetric design of the wire pairs, by careful routing of the common mode path, and by avoiding proximity between sensitive lines. 

The dominant source of nonlinearity in the signal path is the SQUID transfer function. The amount of intermodulation products is a function of coincidence of modulated signals, and their amplitude. The size of the nonlinear effect of the SQUID transfer function can be designed by choosing the maximum flux excursion, at the cost of SQUID dynamic range, and by negative feedback. The requirements for non-linearity are still subject of investigation.  End-to-end simulations are ongoing\cite{denhartog16} to provide this information.

Electromagnetic interference (EMI) creates signals which add to the detector signals, and which impair the energy resolving power when their levels exceed the NEP levels of the detector ($\sim 10^{-18} \mathrm{W}/\sqrt{\mathrm{Hz}}$). Careful shielding, i.e. creating a solid Faraday cage around the detector and the low-level signals, is the best way to keep the EMI levels below the threshold levels. The effects of in-band EMI can further be attenuated by using balanced signals. For out-of-band EMI, further attenuation can be obtained by filtering.

\section{READOUT DESIGN CONCEPT FOR X-IFU}

\begin{table}
\centering
\begin{tabular}{||l|l||}
\hline
{\bf Property} & {\bf Requirement} \\
\hline
\hline
Energy range & 0.2 - 12 keV \\
Energy resolution budget SQUID + LNA  & 2.7 \% of $\Delta$E = 2.5 eV FWHM @ 7~keV \\
Count rate extended sources & $\lesssim 2$~cts/s/pixel  \\
Count rate point sources &  $\lesssim 50$~cts/s/pixel within the PSF \\
Linearity SQUID tandem & THD $<$ 1\% for $E< 2$~keV (TBC) \\
Number of pixels & 3832 \\
Pixel bias power & $< 6$~pW \\
Pixel thermal bandwidth & 0.4~kHz\\
Cross talk &  TBD\% \\
EMI induced noise level after shielding and filtering & 30\% of LNA noise level \\  
\hline
\end{tabular}
\caption{Summary of the main X-IFU requirements and pixel array characteristics with relevance for the design of the readout chain.}
\label{tab_reqs}
\end{table}

\begin{figure}
\centering
\includegraphics[width=0.8\textwidth]{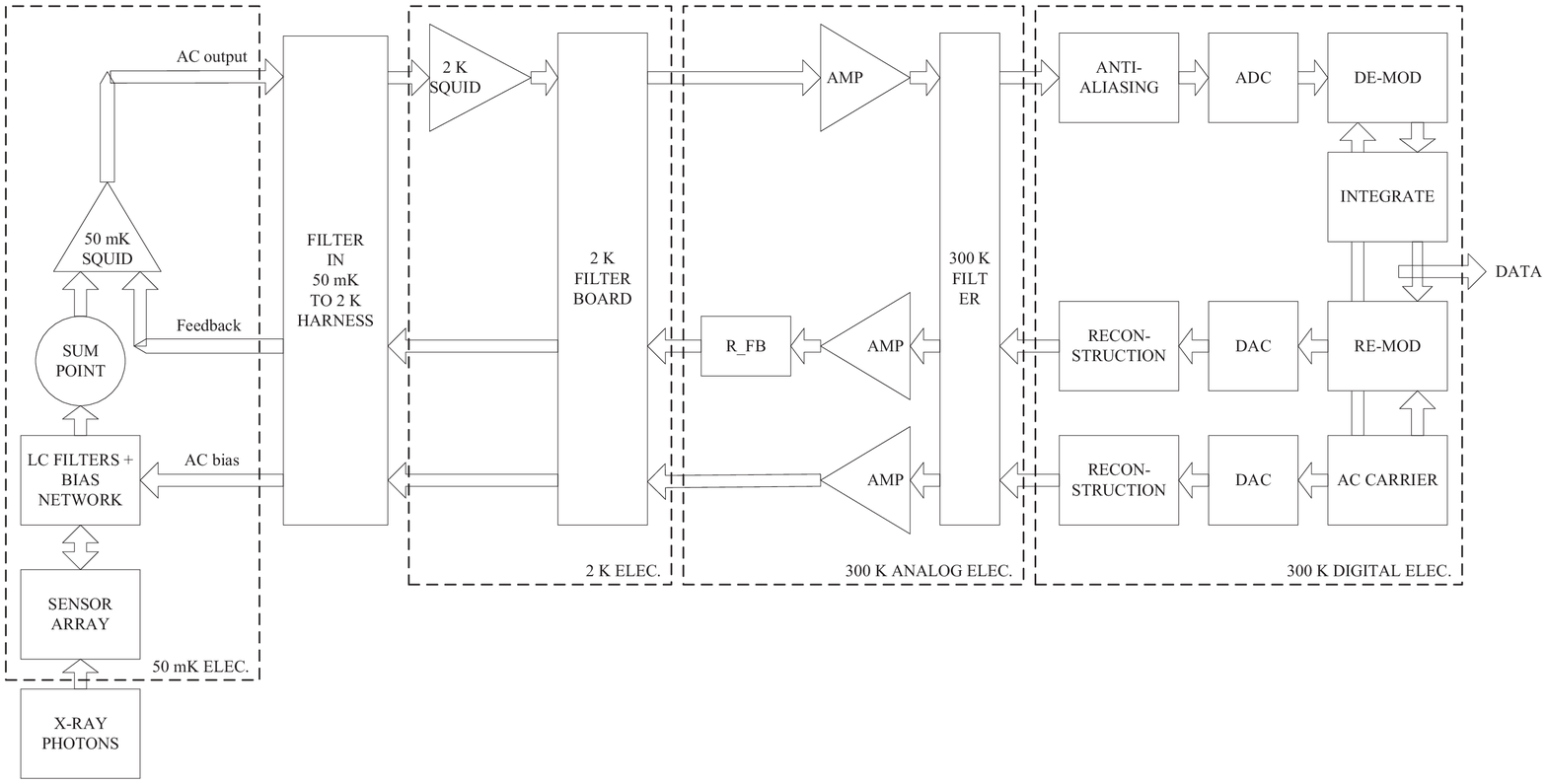}
\caption{overview of the FDM implementation concept for X-IFU. The cold electronics at 50mK consists of the sensor array, $LC$ filters and front end SQUIDs. The electronics at the 2K level consists of the booster SQUIDs and EMI filters. The analog front end electronics at 300K provides signal and buffer amplifiers, as well as bias sources for the SQUIDs, heaters, and temperature sensors. The digital electronics provides the demodulation functionality, the sinusoidal (AC) carrier generation, and feedback generators to enhance the (linear) dynamic range of the  SQUID amplifiers.}
\label{fig_bblocks}
\end{figure}

Fig.~\ref{fig_bblocks} shows an overview of the FDM implementation concept for X-IFU. The cold electronics at 50mK consists of the sensor array, $LC$ filters and front end SQUIDs. The electronics at the 2K level consists of the booster SQUIDs and EMI filters. The analog front end electronics at 300K provides signal and buffer amplifiers, as well as bias sources for the SQUIDs, heaters, and temperature sensors. The digital electronics provides the demodulation functionality, the sinusoidal (AC) carrier generation, and feedback generators to enhance the (linear) dynamic range of the  SQUID amplifiers.

A summary of the X-IFU requirements which drive the design of the readout chain are summarised in table~\ref{tab_reqs}. A number of requirements are still under study, as  the flow down of the science requirements to instrument requirements is still ongoing. 

\subsection{SQUID chain design}

The most important function of the SQUID amplifier is to raise the TES signals above the noise level of the first stage low noise amplifier (LNA) at room temperature, so that the contribution of the room temperature electronics to the energy resolution budget remains small, i.e. $\leq 2.7\%$ following the preliminary energy resolution budget for X-IFU. To achieve this, a minimum amount of power gain in the SQUID is required\cite{kiviranta11}. The amount of required power gain will be estimated below.

The minimum required amount of power gain $G_P$ can be calculated from the noise temperatures a $G_P> (T_A + T_{N,sq})/T_{N,tes}/p_{\mathrm{budget}}/F$,  where $T_{N,tes} \approx T_c=100$mK the noise temperature  of the TES, $p_{\mathrm{budget}}=0.016$ the budget assigned to the SQUID/LNA combination of 2.7\%, and $F$ the mismatch factor between the output dynamic resistance of the SQUID and the noise matching resistance of the LNA.
The noise temperature $T_{N,\mathrm{LNA}}$ of a state-of-the-art LNA which is suitable to approximately noise-match to the low ($\sim 120 \Omega$) output impedance of a SQUID  equals $T_{N,\mathrm{LNA}} \gtrsim 70$K in SiGe based technology.  
The intrinsic noise temperature\cite{kiviranta06} at the output of an optimised SQUID operated at a temperature $T_{sq}$, equals $T_{N,sq} \gtrsim 2.4 T_{sq} $, which implies that the LNA is the dominant noise source. So for X-IFU, where SQUID operation is intended at $T_{sq} \leq 2$~K, we estimate that $P_G > 54$dB for a SQUID with a noise mismatch factor of $F\sim 0.2$.
A SQUID with a relative high critical current density of $J_c = 500$A/cm$^2$, the theoretical maximum power gain per stage equals $\sim 32$~dB @ 5~MHz. So, a single stage SQUID is insufficient to provide the required power gain.

There are different ways in which the power gain can be increased. For example, it has been shown that positive feedback \cite{drung90,kiviranta08} can be used to enhance the power gain moderately, at the cost of bandwidth and enhanced vulnerability to SQUID transfer variations. However, when other requirements such as linearity, stability, power dissipation, and multiplex factor (= bandwidth) are also taken into account, a two-stage SQUID cascade, though more complex, turns out to be more robust and attractive as the baseline for X-IFU. This is caused by the fact that all the required properties of the SQUID, such as the coupled energy resolution to minimise its noise contribution, the output resistance to drive a cable with sufficient bandwidth to enhance the multiplex factor, the power gain to boost the signals above the noise floor of the LNA, and the dynamic range density $D$ to match the signal-to-noise ratio of the micro-calorimeter pixels to the SQUID range, scale linearly with the power dissipation of the SQUID with the exception of the dynamic range density which scales quadratically. 

In the two stage configuration the front end SQUID is located close to the detector at the base temperature of the system. The second stage (``booster'') SQUID resides at the 2K level, where the available cooling power is ample with respect to the base temperature stage, and the proximity to the first stage is much closer than to the LNA. In such a two-stage configuration, not only is sufficient power gain available, but the power dissipation of the front end SQUID can also be much smaller than in a single stage approach.  This originates from two main differences between the two stages. First, the front end SQUID  only needs to drive the (short) interstage cable and the input coil of the booster SQUID, while the booster SQUID drives the longer cable (factor of $\sim 10$) to the LNA. Second, the dynamic range density of the front end SQUID is not limited by the LNA noise temperature but by its own intrinsic flux noise instead, which saves a factor of $\sqrt{T_A /2.4 T_{sq}} \approx 10$ for an optimised SQUID design.

The baselined SQUID tandem consist of a multiloop front end, and a large array SQUID as booster stage. A summary of the properties is presented in table~\ref{tab_squidprops}. A more extended description has been described elsewhere\cite{kiviranta14}.  The output of the booster SQUID is balanced to enhance immunity against common mode disturbances which are picked by the cable between the SQUID and the room temperature electronics. Its signal output power is maximised in order to optimise the net dynamic range density. The front end SQUID is dimensioned such that it can drive the full range of the booster SQUID within the required interstage bandwidth, again to optimise the dynamic range density of the system, and consequently the multiplex factor. Simultaneously, this optimisation sets a lower limit to the power dissipation in the front end SQUID to approximately 2 nW within the chosen parameter space.

\begin{table}
\centering
\begin{tabular}{|l||ll|}
\hline
{\bf Property} & {\bf front end SQUID} & {\bf booster SQUID} \\
\hline
operating temperature & 50mK & 2K \\
input inductance & $<3$nH & $<160$nH \\
power dissipation & $<2$nW & $>0.5 \mu$W \\
input current noise & $<3$ pA/$\sqrt{\mathrm{Hz}}$ & N/A \\
operating mode & single ended & differential \\
bandwidth 2-stage system & \multicolumn{2}{l|}{$>8$MHz}\\
interstage cable inductance & \multicolumn{2}{l|}{$<80$nH}\\
linearity 2-stage system & \multicolumn{2}{l|}{$< 1$\% THD, (TBD)}\\
\hline
\end{tabular}
\caption{Summary of the SQUID properties for the X-IFU readout.}
\label{tab_squidprops}
\end{table}

\subsection{Frequency space assignment}

The available frequency space on the upper side is driven by a number of factors, such as cable losses which are driven by thermal constraints, dissipation in the digital electronics which scales proportional with the operating frequency, and SQUID bandwidth which scales linear with the power consumption. All these factors combined have led to the choice of an upper frequency limit of 5~MHz, based on the currently available information. When more details on the design space of all components become available, it might be possible to push up this value.
 
The available frequency space for multiplexing is set on the lower side by $LC$ filter component size, as their area consumption scales as $1/f^2$. A practical limit of 1~MHz was chosen, based on the achievable density in the existing technology \cite{bruijn14} which provides approximately 3 nF/mm$^2$ for 28nm of a-Si:H. Minimum area consumption is obtained when on average the coil and capacitor size are of similar size. For multiplexing in the MHz region, this has led to the choice for coils with an inductance of 2~$\mu$H.

The frequency separation between pixels is driven by cross talk requirements, and SQUID dynamic range enhancement requirements using baseband feedback. This has led to the choice of 100~kHz separation between adjacent carrier frequencies, which is compatible with a gain-bandwidth product per pixel of $\mathrm{GBP} \sim 15$~kHz. The available frequency space matches the choice for a multiplexing factor of $M=40$, based on the SQUID power dissipation discussion above.

\subsection{Interconnections and transformations}
\label{sec_intercon}

\begin{figure}
\centering
\includegraphics[width=0.5\textwidth]{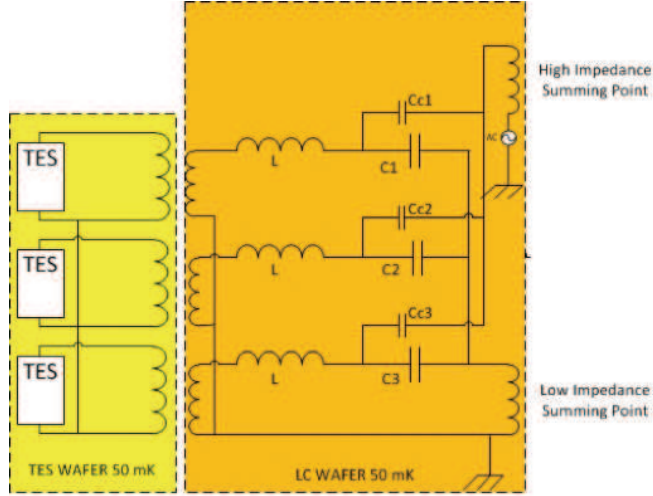}
\caption{Schematic of detector wafer and $LC$-wafer with transformer coupled interconnections for three pixels. The transformation ratio matches the TES resistance to the filter inductance $L$, so that the minimal required bandwidth for electro-thermal TES operation is obtained. There are two summing points indicated in the diagram: The high-impedance summing point in which the voltage source is implemented, and the low-impedance summing point in which the SQUID input coil is located (the SQUID not indicated). The high impedance summing point has been created to up-transform the TES resistance, so that stray inductances and resistances become acceptable in the summing point. In this way, a shared voltage source can be created with minimal parasitic effects.  }
\label{fig_summing}
\end{figure}

The sensor array wafer, the $LC$ filter wafer, and the SQUID wafers are all produced in different technologies and by different consortium members. Consequently, a monolithic approach is impossible, so that interconnections are needed between the different building blocks. From instrument integration and testing point of view, it is preferable to have a configuration in which components can be tested independently before assembly, and in which components can be reworked.

The low TES resistance in combination with the voltage bias condition makes the system sensitive to impedances as a result of stray inductances. In alternating current (AC) systems, impedance transformation is a classical method to transform impedance levels to convenient values.  This method has also been applied in the X-IFU readout design at two locations, as indicated in fig.~\ref{fig_summing}.

 Magnetic transformers are used to provide the impedance matching between the $LC$-filter and the TESs. The transformation ratio matches the TES resistance to the filter inductance $L$ such, that the minimal required bandwidth for electro-thermal TES operation is obtained. 
 
 The magnetic transformers make it possible to avoid wire or bump bonded interconnections, by creating an inductive interconnection by implementing one coil of the transformer on the TES side, and the other side of the transformer on the $LC$ filter side. The mechanical aspects of this transformer are discussed elsewhere\cite{jackson16}. Because of the reworkablity, the transformer coupling has been baselined for X-IFU, with bump bonding as a fall back solution.

There are two summing points indicated in the diagram: the high-impedance summing point in which the voltage source is implemented, and the low-impedance summing point in which the SQUID input coil is located (the SQUID not indicated), including the bonding wires. By creating multiple parallel input coil connections to the SQUID, the stray inductance in the summing point is minimised. 

The high impedance summing point has been created to up-transform the TES resistance, so that stray inductance and resistance levels in the summing point become acceptable. In this way, a shared voltage source can be created with minimal parasitic effects, i.e. common impedance. The high impedance summing point is achieved with capacitive voltage dividers, which up transform the TES resistance with a factor $(Cc_n/C_n)^2$, also provide a part of the range matching between the generated bias voltages (at a volts-level) at room temperature, and the micro-calorimeter bias voltages (at a microvolt level). 

In the baseline design the SQUIDs are located in the low-impedance summing branch because of their low-impedance nature, and the voltage sources in the high-impedance summing branch because of their higher impedance nature. However, there is no fundamental reason why the summing branches couldnÕt be combined, when more information on the properties of a front end SQUID with a higher turn ratio of the input coil is available.

\section{SUMMARY}

The X-IFU instrument on the Athena satellite will provide high resolution ($<$~2.5~eV) imaging X-ray spectroscopy capability in the energy range between 0.2 an 12~keV.  A TES-based micro-calorimeter array will be used in the focal plane to deliver the required combination of properties. The cryogenic operating temperature of $\sim$90~mK requires careful design electrical readout, in order to fit within the physical and thermal boundary conditions as set by the space craft.

The combination of the noise temperature of the micro-calorimeters and the noise temperature of the room temperature amplifier makes SQUID amplifiers the only option to provide  the required power gain. The intrinsic bandwidth of a SQUID-based amplifier with respect to the much smaller bandwidth of a micro-calorimeter pixel makes that there is frequency space available for multiplexing.
Multiplexing is needed to minimise the wire count to practically feasible proportions, and to minimise the heat load at the cryogenic stages. 

Frequency domain multiplexing (FDM) is attractive for the X-IFU readout as it minimises the dynamic range requirements on the SQUIDs, which in turn helps to minimise the heat load of the SQUIDs. A multiplex factor of 40 has been baselined, as a result of a compromise between available bandwidth, required pixel speed, physical area consumption, and cross talk requirements. 

A two stage SQUID amplifier is needed to deliver a combination of sufficient dynamic range density (driven by the signal-to-noise ratio) and power gain. The first stage operates at the base temperature, while the booster SQUID operates at 2K. The power dissipation is driven by the dynamic range and bandwidth requirements. Further analysis of the cross talk requirements is needed to further optimise the SQUID chain.   

Transformer-based interconnections will provide a combination of a reworkable focal plane assembly, and optimal impedance matching between the TESs and the $LC$ bandpass filters. Capacitive dividers further help to engineer the parasitic effects, which are the result of the low ($\sim$m$\Omega$) operating resistances of the TESs.  

\acknowledgements

The research leading to these results has received funding from the European UnionÕs Horizon 2020 Programme under the AHEAD project (grant agreement n. 654215).


\end{document}